# Some consequences of a Universal Tension arising from Dark Energy for structures from Atomic Nuclei to Galaxy Clusters

*C Sivaram*

Indian Institute of Astrophysics, Bangalore, 560 034, India

Telephone: +91-80-2553 0672; Fax: +91-80-2553 4043

e-mail: sivaram@iiap.res.in

*Kenath Arun*

Christ Junior College, Bangalore, 560 029, India

Telephone: +91-80-4012 9292; Fax: +91-80- 4012 9222

e-mail: kenath.arun@cjc.christcollege.edu

*Kiren O V*

Christ Junior College, Bangalore, 560 029, India

Telephone: +91-80-4012 9292; Fax: +91-80- 4012 9222

e-mail: kiren.ov@cjc.christcollege.edu

**Abstract:** In recent work, a new cosmological paradigm implied a mass-radius relation, suggesting a universal tension related to the background dark energy (cosmological constant), leading to an energy per unit area that holds for structures from atomic nuclei to clusters of galaxies. Here we explore some of the consequences that arise from such a universal tension.

## 1. Introduction

In recent papers (Sivaram, 1994a; 1994b; 2008; Sivaram & Arun, 2012a; 2013) a new kind of cosmological paradigm was invoked wherein the requirement that for a hierarchy of large scale structures, like galaxies, galaxy clusters, super-clusters, etc. their gravitational (binding) self energy density must at least equal or exceed the background repulsive dark energy density (a cosmological constant as current observations strongly suggests) implies a mass-radius relation of the type:

$$\frac{M}{R^2} = \frac{c^2}{G}\sqrt{\Lambda} \qquad \ldots (1)$$

(i.e. $\frac{GM^2}{8\pi R^4} = \frac{\Lambda c^4}{8\pi G}$ gives rise to a universal tension, $T = \frac{Mc^2}{R^2} = \frac{c^4}{G}\sqrt{\Lambda}$)

This is the background curvature x superstring tension (Sivaram, 1994a; 2005). Or this can also be inferred as the local mass x local curvature. (Superstring tension is $\sim \beta c^4/G$, $\Lambda$ is the cosmological dark energy)

This paradigm focuses on the universe's fundamental structures and symmetries and emphasises a new universal parameter underlying systems from the smallest (atomic nuclei) to the largest (clusters of galaxies), encompassing nearly 80 orders of magnitude in mass and nearly 40 orders of magnitude in size. (Oldershaw, 1987; Sivaram, 1993; 2001; 2005)

## 2. Nuclear Tension

The energy per unit area (surface tension) given by above equations, i.e. $T = \frac{Mc^2}{R^2} = \frac{c^4}{G}\sqrt{\Lambda}$, has the same numerical value as that in nuclear physics, (Sivaram, 2005; 2008) like the surface tension in the nuclear liquid drop model of $\sim 10^{21} ergs/cm^2$. This has consequences for the nucleus and nuclear matter.

In the nucleus this nuclear surface tension balances the Coulomb repulsion:

$$\frac{Z^2 e^2}{R} = 4\pi R^2 T \qquad \ldots (2)$$

Where $R = R_0 A^{1/3}$, $R_0 \sim 1.5 \times 10^{-13} cm$

For $T \sim 10^{21} ergs/cm^2$, this sets a limit of:

$$\frac{Z^2}{A} < 40 \qquad \text{... (3)}$$

which agrees with the usual Bohr-Wheeler criterion. (Bohr & Wheeler, 1939)

Considering also the rotation (spin) of the nucleus we have:

$$\frac{Z^2 e^2}{R} + \frac{1}{5} M \omega^2 R^2 = 4\pi R^2 T \qquad \text{... (4)}$$

Where the terms on the left are repulsive in nature and that on the right is attractive.

This gives the radius of the size of the nucleus as (where $M = m_P A$, $m_P$ is the proton mass):

$$R = \left( \frac{Z^2 e^2}{4\pi T - \frac{1}{5} m_P A \omega^2} \right)^{1/3} \qquad \text{... (5)}$$

The angular frequency dependence on nuclear size is given in figure (1):

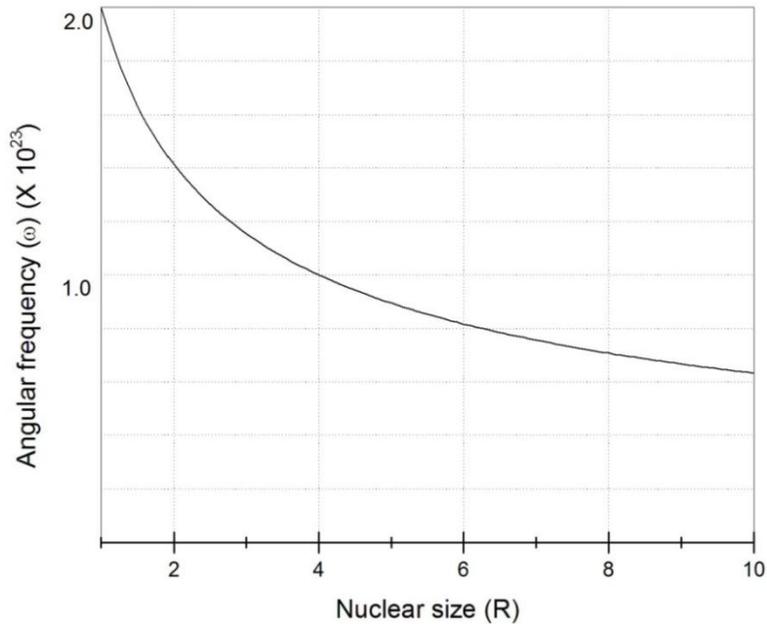

**Figure 1: Dependence of Angular Frequency with Nuclear size**

For the limiting case as $R \to \infty$:

$$4\pi T = \frac{1}{5} m_P A \omega^2 \qquad \ldots (6)$$

This sets a limit on the frequency as:

$$\omega = \left(\frac{20\pi T}{A m_P}\right)^{1/2} \qquad \ldots (7)$$

For $A = 1$ we have:

$$\omega \leq 2 \times 10^{23}\, s^{-1} \qquad \ldots (8)$$

And the corresponding time period of:

$$\tau = \frac{2\pi}{\omega} \sim 3 \times 10^{-23}\, s \qquad \ldots (9)$$

This corresponds to the nuclear time scale. Figure (2) gives the variation of angular frequency with the number of nucleons.

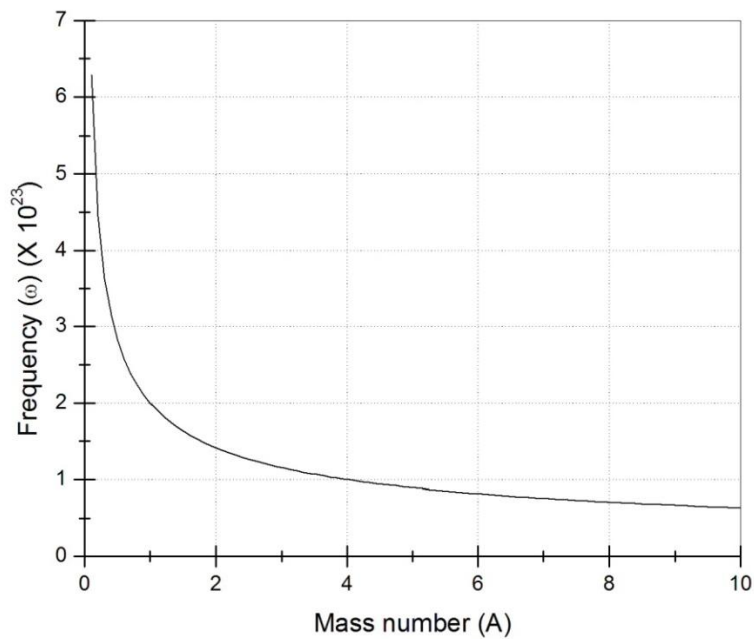

**Figure 2: Variation of angular frequency with A**

This limit on the frequency will also put a constraint on the rotational energy levels of nucleus:

$$\frac{2}{5} M\omega^2 R^3 = n\hbar \qquad \ldots (10)$$

$$n^2 \leq 10 A^{7/3} \qquad \ldots (11)$$

For *A = 10*, we have $n \approx 100$

These states correspond to the yrast states, which are the lowest excited level at high angular momentum $(\sim 70\hbar)$ as suggested in the following reference (Grover, 1967). Later observations indicate high angular momentum $(\sim 100\hbar)$.

Figure (3) gives the almost linear dependence of the order (*n*) with the mass number.

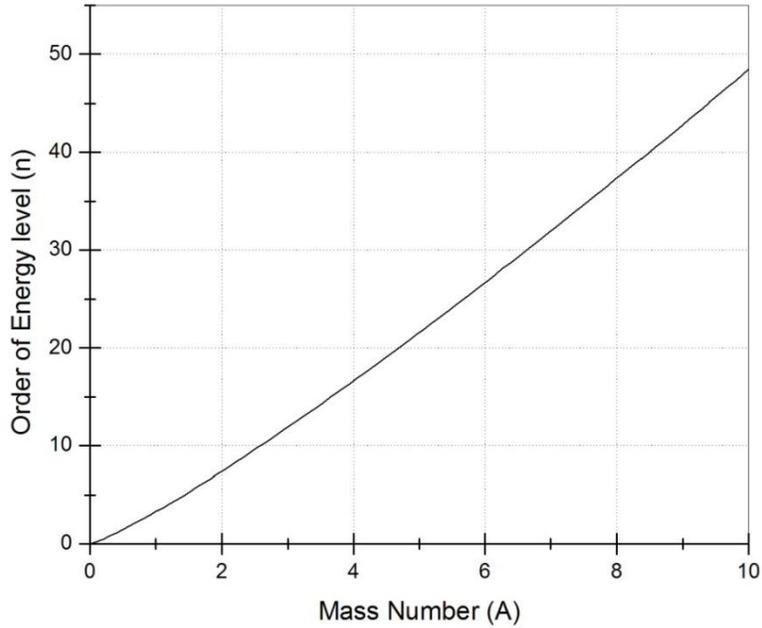

**Figure 3: Variation of *n* with Mass number**

## 3. Gravitationally bound structure, Angular momentum and Dark Energy

In the case of large, gravitationally bound structures such as galaxies, galaxy clusters, etc. the requirement is that gravitational self energy density should be comparable to the background cosmic vacuum energy density for the object to be an autonomous structure. That is:

$$\frac{GM^2}{8\pi R^4} = \frac{\Lambda c^4}{8\pi G} \qquad \ldots (12)$$

This would also give the same result as equation (1), i.e.: $\frac{M}{R^2} = \frac{c^2}{G}\sqrt{\Lambda}$

Where $\Lambda$ is the cosmological constant with an observed value of $10^{-56} cm^{-2}$. This equation holds for a whole range of large scale structures, including the Hubble volume. (Sivaram, 2008 and references there in)

The $\frac{M}{R^2}$ relation is suggestive of a surface tension which has the same universal value for all the large scale cosmic structures from globular clusters, large molecular clouds, all the way to the Hubble universe (Sivaram, 1994a; 2008). A kind of universal surface tension, suggesting the holographic picture! (Sivaram & Arun, 2013) (It also holds for the electron!)

The universality of this surface tension again constrains the size of a neutron star. For a neutron star composed on 'N' neutrons:

$$(4\pi R_{NS}^2 T)N = \frac{GM_{NS}^2}{R_{NS}} \quad \text{... (13)}$$

For a 2 solar mass neutron star, $N \approx 5 \times 10^{57}$, which matches the observations for the heaviest detected neutron star. (Crawford, et al, 2006)

Considering also the rotation of the neutron stars we have:

$$(4\pi R_{NS}^2 T)N = \frac{GM_{NS}^2}{R_{NS}} - M_{NS}\omega^2 R_{NS}^2 \quad \text{... (14)}$$

This sets a limit on the rotational frequency and the corresponding time period of the neutron star as:

$$\omega \approx 10^4 s^{-1}, \ \tau \approx 0.5 ms \quad \text{... (15)}$$

This is consistent with the observations of the millisecond pulsar having the fastest rotational period detected so far, which is ~1.3ms. (Hessels, et al, 2006)

In the case of galaxies, this surface tension balancing the rotational energy can possibly explain the flat rotation curve of the galaxies. That is, for galaxies, their rotation balances this surface tension. This gives:

$$4\pi R_{gal}^2 T = M_{gal}\omega^2 R_{gal}^2 \quad \text{... (16)}$$

Where the rotation frequency and the corresponding time periods are given as:

$$\omega = \left(\frac{4\pi T}{M}\right)^{1/2} \approx 3 \times 10^{-12} s^{-1}, \quad \tau \approx 10^{12} s \qquad \ldots (17)$$

Since $M/R^2$ is a constant even for a galaxy, the relation given by equation (16) leads to:

$$\omega \propto \frac{1}{R_{gal}} \qquad \ldots (18)$$

$$\omega R_{gal} = \text{constant}$$

Therefore, the velocity, which is given by:

$$v = \omega R_{gal} = \left(\frac{4\pi T}{M_{gal}/R_{gal}^2}\right)^{1/2} \qquad \ldots (19)$$

will also be a constant as expected from the galaxy rotation curves! This suggests a velocity independent of radial distance (flat rotation curve) without invoking dark matter.

It is interesting to note that this dependence of rotational frequency going as inverse of the size hold true even right down to the atomic nucleus, as indicated by figure 4.

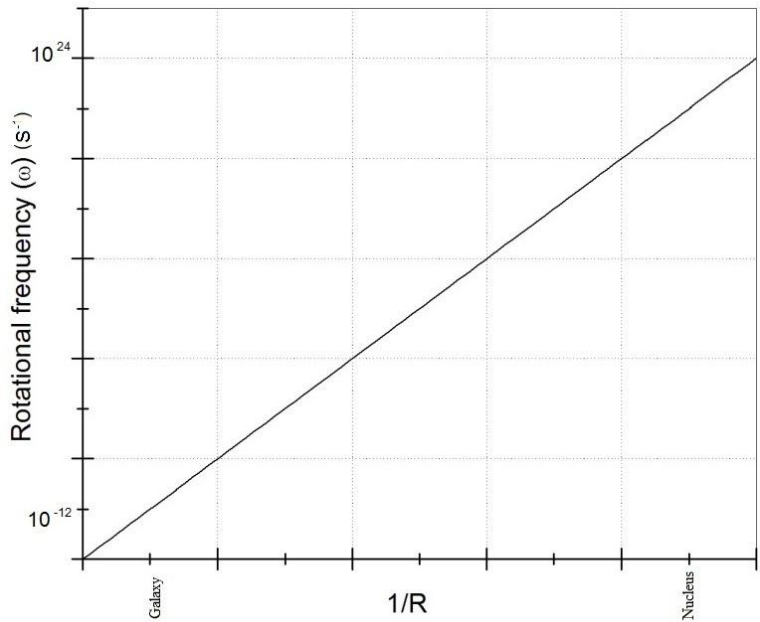

**Figure 4: Variation of rotational frequency with size**

In an earlier work (Sivaram & Arun, 2012b) a primordial cosmic rotation was suggested which can give rise to the observed rotation angular momenta of galaxies, galaxy clusters, stellar planetary systems, etc., the origin of which is otherwise not clearly understood.

The angular momentum of the galaxy given by $J_{gal} = M_{gal}\omega R_{gal}^2$, is conserved. Therefore we have (again since $M/R^2$ is a constant):

$$\frac{J_{gal}}{R_{gal}^3} = \frac{M_{gal}\omega R_{gal}^2}{R_{gal}^3} = \frac{M_{gal} v R_{gal}}{R_{gal}^3} \text{ is a constant} \qquad \text{... (20)}$$

This also leads to $\omega R_{gal} = \text{constant}$ and therefore a velocity independent of radial distance.

## 4. Dynamics of evolving structures

The requirement that the gravitational self energy density must at least equal or exceed the background repulsive dark energy density implied a mass-radius relation as given by equation (1), for a hierarchy of large scale structures, like galaxies, galaxy clusters, super-clusters, etc.

This mass-radius relation holds good for nebulae too. Any perturbation to it will lead to its collapse and eventual formation of the star (and possibly planetary system). For a typical star of mass, $M_{star} \approx 10^{33} g$, the condition that $M/R^2 \approx 1$ implies that the initial size of the nebula be of the order of $R_{Neb} \sim 3 \times 10^{16} cm$.

It is also of interest to note that the same value for the tension (arising as we have seen, from the cosmic dark energy ($\Lambda$ term)) which we have used for galaxies, galaxy clusters, atomic nuclei, etc. also seems to be relevant for the dimensions of planets and stars.

For example, for a typical planetary mass of $M \sim 10^{28} g$, balancing surface energy and gravitational self energy, i.e.

$$4\pi R^2 T = \frac{GM^2}{R} \qquad \text{... (21)}$$

we get the radius, which is given by:

$$R = \left(\frac{GM^2}{4\pi T}\right)^{1/3} \qquad \ldots (22)$$

For $M \sim 10^{28} g$, we get $R \approx 5000 km$ (the earth radius). The above equation also gives a Jupiter radius of $\sim 10^5 km$ for the corresponding mass.

For a typical stellar mass of $M \sim 10^{33} g$, the above equation implies $R \sim 10^{12} cm$. So the range of stellar and planetary sizes is also given by the same value of *T*! This suggests a deep underlying connection between the background dark energy (Λ-term, which gives the background curvature) and all the structures embedded in this background. For the large structures we had balance of gravitational energy densities with the background dark energy density. For the planetary and stellar objects, the balance was with surface energies and gravitational self energies.

## 5. Densities of various structures

As noted above, we had a universal, $M/R^2$ ratio, i.e. a ubiquitous surface tension of $\sqrt{\Lambda}\,c^4/G \approx 10^{21} ergs/cm^2$, underlying all entities from nuclei to galaxy superclusters! But we know that nuclear density is $\sim 10^{14} g/cc$, superclusters have a density of $\sim 10^{-25} g/cc$. How to understand this diversity in densities?

It is just that the average density is $\sim \dfrac{M}{R^3}$, so that if we have the universal $T = \dfrac{M}{R^2} = \dfrac{c^2}{G}\sqrt{\Lambda} \sim 1 g/cm^2$, the densities of the various structures considered would scale as $\dfrac{M}{R^3}$, i.e. $\rho \propto \dfrac{T}{R}$! (As $T = M/R^2$, so $M/R^3$ is just $T/R$)

As *T* is a universal constant the density simply scales as $1/R$. (In connection with surface tension, this is the Laplace pressure for a droplet).

Thus for a nucleus $R \sim 1\,fermi$, we have $\rho \sim 10^{13} - 10^{14}\,g/cc$. For a galaxy $R \sim 10^{23}\,cm$, we have $\rho \sim 10^{-23}\,g/cc$. For a super-cluster $R \sim 10^{25}\,cm$, $\rho \sim 10^{-25}\,g/cc$. And for the Hubble volume, $\rho \sim \dfrac{1}{R_H} \sim \dfrac{1}{\sqrt{\Lambda}} \sim 10^{-29}\,g/cc$, just what is observed!

So we have another universal result:

$\rho R = $ constant  ... (23)

Holding from nuclei to the universe!

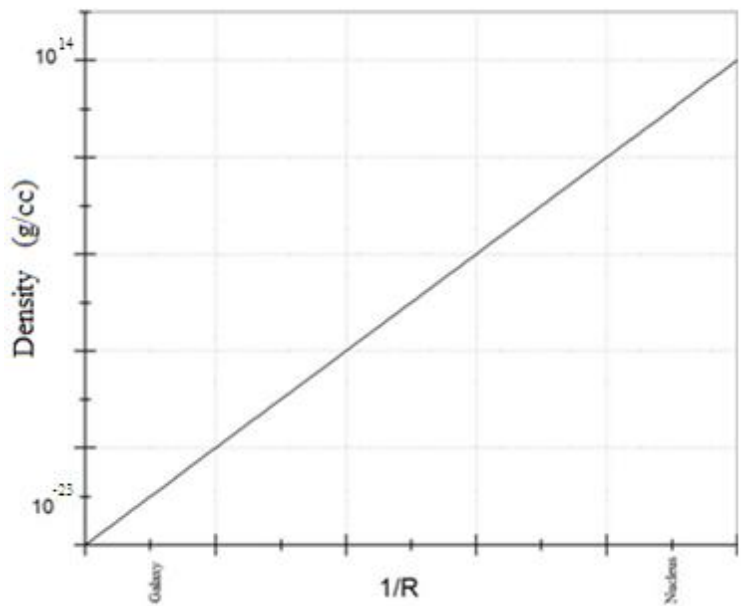

**Figure 5: Variation of density with size**

## 6. Nuclear Vibrational and Rotational Energy levels

In connection with the energy levels of the nucleus, including both vibrational and rotational levels, we can invoke the liquid drop model of nuclei. In the drop model there is equilibrium between surface tension and Coulomb repulsion. Small perturbations of the drop surface of radius R by $\delta(r)$ gives changes in surface energy (surface given by $F(r,\theta,\phi) =$ constant), which can be expanded in spherical harmonics (like in fluid mechanics of incompressible liquid spheres).

Thus:

$$\delta = R\sum_{l,m} C_{l,m} Y_{l,m} \qquad \ldots (24)$$

$$L^2 Y_{l,m} = l(l+1) Y_{l,m} \qquad \ldots (25)$$

The surface energy is perturbed as:

$$\delta E_S = \frac{TR^2}{2} \sum_{l,m} (l-1)(l+2) C_{l,m} C_{l,m}^* \qquad \ldots (26)$$

While the electrostatic (repulsive) Coulomb energy is perturbed as:

$$\delta E_C = -\frac{3Z^2 e^2 (l-1)}{4\pi(2l+1)} C_{l,m} C_{l,m}^* \qquad \ldots (27)$$

Finally we can write the Hamiltonian including also the kinetic energy:

$$\sum_{l,m} \frac{\rho R^5 \omega^2}{2l} C_{l,m} C_{l,m}^* \qquad \ldots (28)$$

The lowest mode being $l = 2$ we have the energy levels of a five-dimensional harmonic oscillator as:

$$E = \hbar \omega_l \left(n + \frac{5}{2}\right) \qquad \ldots (29)$$

This for $l = 2$ gives the ground state energy level as:

$$E_0 = \frac{5}{2} \hbar \left( \frac{8\pi}{3Am_n} \left( 4T - \frac{3Z^2 e^2}{10\pi R^3} \right) \right)^{1/2} \qquad \ldots (30)$$

For a nuclei of $Z = 20$, $A = 40$, the above equation gives a ground state energy of:

$$E_0 \approx 10^{-5} ergs \approx 10 MeV \qquad \ldots (31)$$

The higher levels will be in multiples of $10 MeV$.

For $l = 0$, stability is given by:

$$T < \frac{3Z^2 e^2}{4\pi R^3}, \quad R = R_0 A^{1/3} \qquad \ldots (32)$$

The higher vibrational excited states are given by $n = 1, 2, \ldots$, etc.

We can include the rotational energy levels (like in atomic spectroscopy). Thus rotational levels are $n\hbar\omega_{rot}$. The limiting values of $\omega_{rot}$ for various $A$ have been given above.

Energy levels of rotation are:

$$E_{rot} = \frac{\hbar^2}{2I} l(l+1) \qquad \ldots (33)$$

Where the moment of inertia of the nuclei is: $I = \frac{2}{5} MR^2$

The rotational energy is then given as:

$$E_{rot} \approx 0.1 MeV \qquad \ldots (34)$$

And the total energy is:

$$E_{total} = E_{vib} + E_{rot} \qquad \ldots (35)$$

For various $n$, $l$, etc.

Similar relations as those above hold also for (nuclei of) primordial galaxies, provided we replace the Coulomb energy term with the gravitational energy. This would also have a negative sign as it is binding energy.

In other words the replacement $\left(\frac{Ze}{R^3}\right)^2$ by $G\rho^2$ would give the result. That is, the surface energy is perturbed as:

$$\delta E_S = \frac{TR^2}{2} \sum_{l,m} (l-1)(l+2) C_{l,m} C_{l,m}^* \qquad \ldots (36)$$

And the gravitational energy is perturbed as: (Lamb, 1945)

$$\delta E_G = -\frac{3G\rho^2 R^6 (l-1)}{4\pi(2l+1)} C_{l,m} C_{l,m}^* \qquad \ldots (37)$$

The tension ($T$) term would be the same. Scaling relations are as before and equation (30) will not apply to galaxies! The frequency of oscillation due to the perturbation for the galaxies is given as:

$$\omega = \left(\frac{8\pi}{3}\left(\frac{4T}{M} - G\rho\right)\right)^{1/2} \qquad \ldots (38)$$

For a typical galaxy of $M \approx 10^{44} g$, $\rho \approx 10^{-24} g/cc$, the frequency is $\omega \approx 10^{-11} s^{-1}$.

These oscillations will emit gravitational waves, where the quadrupole gravitational power is given by: (Sivaram & Arun, 2011)

$$P_{GW} = \frac{G}{c^5} M^2 R^4 \omega^6 \qquad \text{... (39)}$$

And for a typical galaxy of $M \approx 10^{44} g$, $R \approx 10^{23} cm$, this gives:

$$P_{GW} \sim 10^{54} ergs/s \qquad \text{... (40)}$$

The corresponding strain produce on a detector, which is given as $h = \frac{GE_{GW}}{c^4 r} \sim 10^{-20}$, which is within the limits of proposed space based gravitational wave observatories like LISA.

## 6. Conclusion

In this paper, we have extended our earlier work (which gave rise to a mass-radius relation) with a universal value of a surface tension $(\sim 10^{21} ergs/cm^2)$ arising from the requirement that the binding energy density of gravitationally bound objects be at least equal or exceed the background repulsive dark energy density. This universal tension arising from dark energy dominating three-fourths of the universe, leads to various consequences for a hierarchy of objects, from atomic nuclei to galaxy clusters. This can for instance set a limit on the rotational energy levels of a nucleus; set the dimensions of planets and stars; to even explain the flat rotation curve of galaxies without invoking dark matter and limit the size of galaxy clusters. In short, we have a new paradigm encompassing features of structures ranging over eighty orders in mass and forty orders in length scale.